\documentclass[
  draft            
  ]
  {aipproc}

\layoutstyle{6x9}

\begin{document}

\title{Radiative decays with $a_0(980)$ and $f_0(980)$ from ChPT at order $p^4$}


\author{S.~Ivashyn}{
  address={
  NSC ``Kharkov Institute for Physics and Technology'',
  Institute for Theoretical Physics,\\
  1, Akademicheskaya str., Kharkov 61108, Ukraine
  }
}

\author{A.~Korchin}{
  address={
  NSC ``Kharkov Institute for Physics and Technology'',
  Institute for Theoretical Physics,\\
  1, Akademicheskaya str., Kharkov 61108, Ukraine
  }
}

\date{May 27, 2008}

\begin{abstract}
\footnote{
To appear in the proceedings of Workshop on Scalar Mesons and Related Topics 
Honoring 70th Birthday of Michael Scadron (SCADRON 70), Lisbon, Portugal, 11-16 Feb 2008. 
}
A consistent description of the $\pi^0\pi^0$ invariant mass distribution
in the $\phi(1020) \to f_0(980)\gamma \to \pi^0\pi^0\gamma$ decay and the
$\pi^0\eta$ in $\phi(1020) \to a_0(980)\gamma \to \pi^0\eta\gamma$ is
suggested. A search for the consequences of the flavor $SU(3)$ symmetry
for the scalar mesons can be based on such an analysis. In order to
accurately treat the pseudoscalar meson dynamics, which is very important
for the scalar meson decays, we employ Resonance Chiral Theory.
\end{abstract}

\maketitle

\section{Introduction}

This paper focuses on the isotriplet $a_0(980)$ and isosinglet $f_0(980)$
scalar mesons ($S$). These particles, probably together with the
controversial $\sigma$ meson, represent the lightest members of the
scalar meson spectrum. To test this assumption it is important to search
for consequences of the flavor $SU(3)$ symmetry for these particles in
nature. Among different observables, the invariant mass distributions of
$\pi\pi$ and $\pi \eta$ pairs in the $ \phi(1020) \to \pi \pi \gamma $
(or $\pi \eta \gamma $) decays are of considerable interest, because the
$a_0(980)$ and $f_0(980)$ are important intermediate states in these
decays and thereby show up in the spectra. The $e^+ e^-$ experiments in
Novosibirsk~\cite{Aulchenko:1998xy,Achasov:2000ym,Achasov:1998cc,Akhmetshin:1999di,Achasov:2000ku}
and
Frascati~\cite{Aloisio:2002bt,Aloisio:2002bsa,Ambrosino:2006hb,:2007mq,:2007ya}
allow one to study the $\phi$ decays. From the point of view of data
extraction, the neutral final states are preferable since in the decays
$\phi(1020) \to \pi^0\pi^0\gamma,\; \pi^0\eta\gamma$ the photon may be
radiated only from the final state. Therefore the above processes are
rather suitable for study of chiral dynamics.

\begin{figure}
\label{fig:0}
  \includegraphics[height=.15\textwidth]{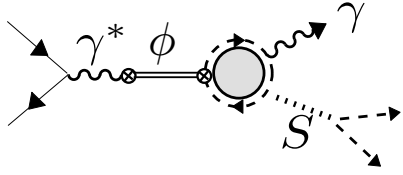}
  \caption{The $e^+e^-$ annihilation to $\pi\pi\gamma $ (and $\pi\eta\gamma$).
  $S$ denotes the intermediate scalar resonance.}
\end{figure}

The analysis of the data (see Refs.~\cite{Achasov:2000ym,Achasov:2005hm} and others)
shows that
  \begin{itemize}
  \item
the decays are mediated by the $S$ resonances, $f_0(980)$ and $a_0(980)$,
in the isoscalar and isovector channels, respectively;
  \item
the kaon loop (KL) coupling of $S$ to the vector meson $\phi$ is very important.
  \end{itemize}
Fig.\ref{fig:0} shows schematically the processes in question. The KL
coupling scheme allows one to reproduce a cusp behavior of the amplitude
dependence on the invariant energy~\cite{Flatte:1976xu}. Many authors
relate the dominance of the KL mechanism to a large $K\bar{K}$ component
in the $a_0(980)$ and $f_0(980)$ mesons and to the proximity of the
$K\bar{K}$ threshold to the scalar meson mass. We argue that the KL
mechanism is a feature of the chiral dynamics and reflects the important
role of pseudoscalar mesons in low- and intermediate-energy interactions.
We also emphasize that the kaon loops in the $\phi \to S \gamma $
transitions naturally arise in the leading order in Resonance Chiral
Theory ($R\chi T$), irrespectively of the threshold or mass position, and
the structure of scalars. In addition, $R\chi T$ accounts for an
important feature -- momentum dependence of the vertex functions.

We have recently obtained~\cite{Ivashyn:2007yy} a complete set of
$\mathcal{O}(p^4)$ contributions to various radiative decays with the
scalar mesons from the $\mathcal{O}(p^4)$ $R\chi T$ Lagrangian. The
parameters were fixed from the average values of the decay widths. In the
following we discuss the $R\chi T$-based study of the $\pi^0\pi^0$ and
$\pi^0\eta$ pair mass distributions in the $\phi(1020) \to S\gamma$
decays and the assumptions used. We fix the parameters and compare
predictions with the data and other models.

We conclude that the invariant-mass distribution is a more reliable
source of information than the average width from PDG, because the width
values may depend on models used in the data analysis.

\section{Formalism}

The $R\chi T$ Lagrangian terms~\cite{EckerNP321}
    \begin{eqnarray}
    L_{scalar}&=&c_d
    \left\langle S^{oct} u_\mu u^\mu \right\rangle + c_m \left\langle S^{oct}
    \chi_+ \right\rangle
+ \tilde{c}_d S^{sing} \left\langle u_\mu u^\mu \right\rangle
    +  \tilde{c}_m S^{sing} \left\langle  \chi_+ \right\rangle
    ,
\\
L_{vector} &=
& \frac{F_{V}}{2\sqrt{2}}\left\langle V_{\mu \nu }f_{+}^{\mu \nu
}\right\rangle + \frac{ iG_{V}}{\sqrt{2}}\left\langle V_{\mu \nu }u^{\mu }u^{\nu
}\right\rangle
    \end{eqnarray}
describe interaction of the scalar singlet $S^{sing}$, octet $S^{oct}$,
pseudoscalar mesons parameterized by $u_\mu$, vector mesons $V_{\mu \nu
}$ and the external electro-magnetic field at order $\mathcal{O}(p^4)$
(for notation we refer to the original paper~\cite{EckerNP321}). The
couplings for the scalar singlet $\tilde{c}_d$ and $\tilde{c}_m$ are in
general independent of the octet couplings ${c}_d$ and ${c}_m$. The
values of $F_V$ and $G_V$ are accurately determined from different
observables. The flavor $SU(3)$ symmetry is explicitly broken by the mass
term contained in $\chi_+$ and the relevant couplings have the subscript
$m$: ${c}_m$ and $\tilde{c}_m$. We mention below the important aspects of
the model~\cite{Ivashyn:2007yy}.
  \begin{itemize}
  \item
In order to estimate the $a_0 \to \pi\eta$ decay width, one has to
construct the physical $\eta$ state from the pseudoscalar singlet
$\eta_1$ and the eighth component $\eta_8$ of the octet. We use the
two-parameter mixing scheme: $\eta = \eta_8 \cos\theta_8  - \eta_1
\sin\theta_1$ with $\theta_8 = -9.2^\circ$ and
$\theta_1=-21.2^\circ$~\cite{Feldmann}.
  \item
For the isoscalar state $f_0 (980)$ we use mixing with the angle $\theta$:
\  $f_0 = S^{sing}\, \cos \theta - S_8^{oct}\, \sin \theta$,
and for the isovector state $a_0 (980)$ we take $a_0 = S_3^{oct}$.
This scheme is rather flexible and allows us to study the mixing within
the lightest nonet.
  \item
Although the usual large $N_c$ behavior of a resonance may not hold for
some physical scalar mesons (see discussion in
Refs.~\cite{Pelaez:2003dy,Jaffe:2007id}), \ the scalars $S^{sing}$ and
$S^{oct}$ in the present model obey this behavior, as these resonances
are introduced similarly to the vector and axial-vector resonances. Thus
we employ the relations $\tilde{c}_m~=~{c_m}/{\sqrt{3}}$ and
$\tilde{c}_d~=~{c_d}/{\sqrt{3}}$ between the singlet and octet couplings.
  \item
The effect of the resonance finite width in the invariant mass
distributions is important~\cite{Oller:2002na}. We choose the propagator
of the scalar meson with the mass $m_{S}$ in the form
    \begin{eqnarray}
\label{scalar-propagator}
D_{S}(p^2)&=&
[p^2 - m_S^2 + i\, m_S \,\Gamma_{S,\;{tot}}(p^2)]^{-1},
    \end{eqnarray}
where the total width $\Gamma_{S,\;{tot}}(p^2)$ depends on the meson
momentum squared $p^2$. It may also be important to use a more advanced
form of the propagator including both real and imaginary parts of the
self-energy, as suggested in Ref.~\cite{Achasov:2004uq}.
  \end{itemize}
Following approach of Ref.~\cite{Ivashyn:2007yy} one arrives at the
matrix elements for the $\phi$ radiative decays.

\section{Discussion}

\begin{figure}
\label{fig:1}
  \includegraphics[height=.4\textwidth]{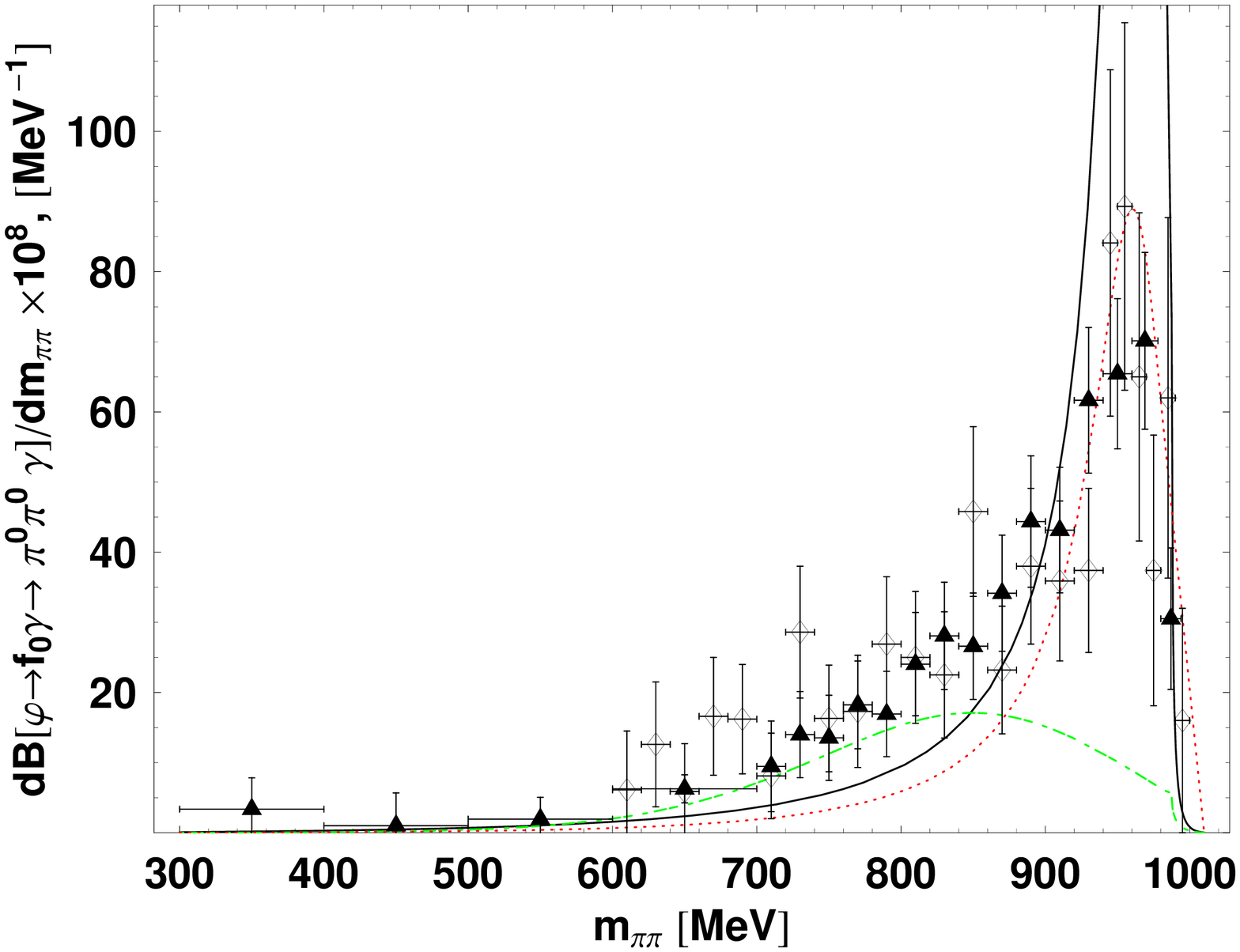}
  \includegraphics[height=.4\textwidth]{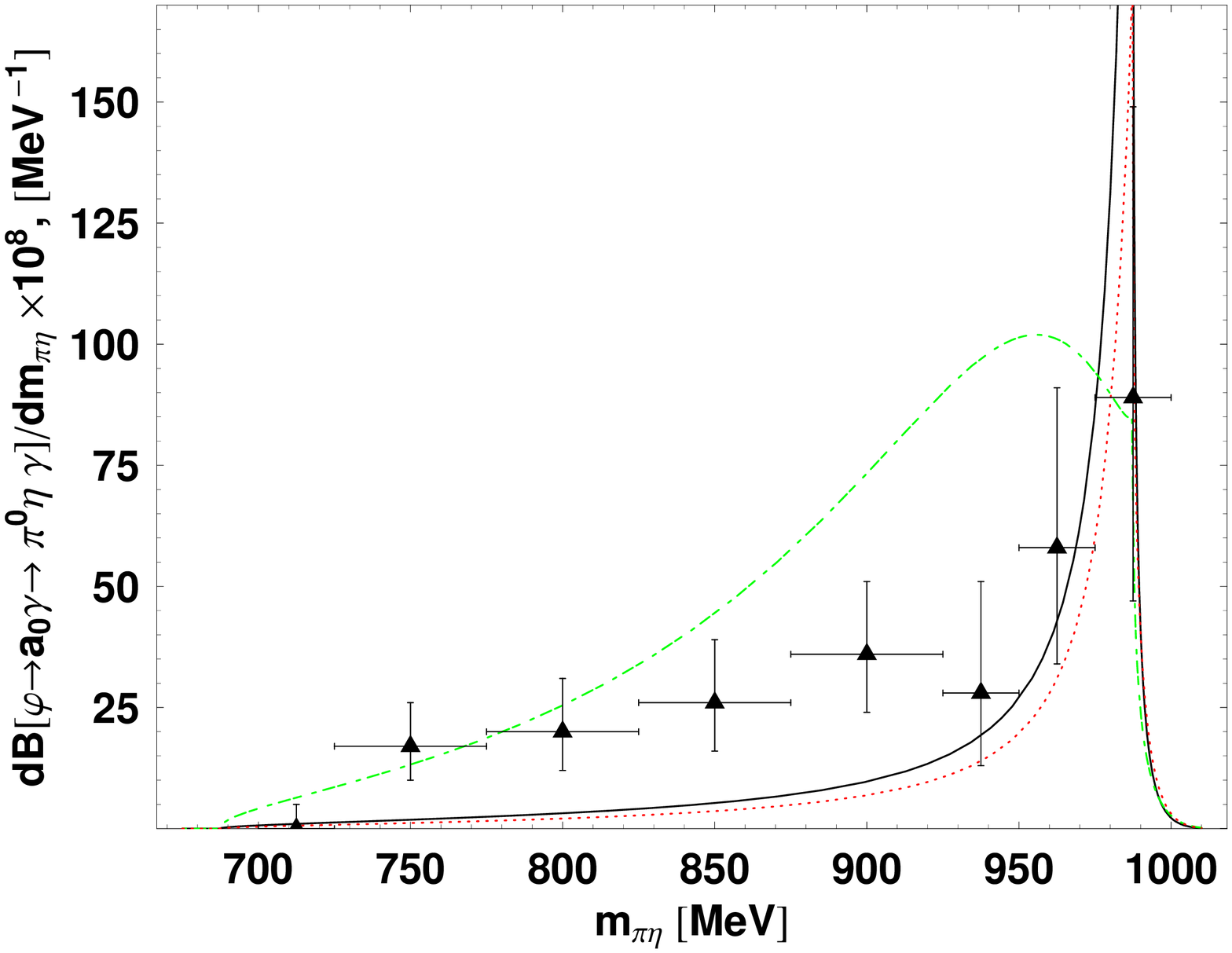}
  \caption{The invariant mass distributions in $\phi\to \pi^o\pi^o\gamma$ ({\it left})
and $\phi\to \pi^o\eta\gamma$ ({\it right}). The solid lines correspond
to the choice of parameters in~\cite{Ivashyn:2007yy}. The dashed lines
are calculated with ``classical'' values for the scalar meson
couplings~\cite{EckerNP321} ($c_m = \;42$~MeV, $c_d = \;32$~MeV, $\theta
= -35.26^o$). The dotted lines are calculated in the present work.
Data:~\cite{Achasov:2000ym}~(left) and~\cite{Achasov:2000ku}~(right). }
\end{figure}

Let us summarize advantages of the present approach.
  \begin{itemize}
  \item
There are no contact $V\gamma S$ and $\gamma\gamma S$ couplings at the
lowest (tree-level) order. Therefore the leading contribution includes
one-loop diagrams with intermediate pseudoscalar mesons. This holds for
any scalar meson mass irrespectively of their internal structure.
  \item
The sum of the loop contributions is convergent, gauge invariant and
universal: one can use either the analytical expression
of~\cite{Close93}, or calculate it numerically.
  \end{itemize}
One should also stress an influence of the pseudoscalar meson dynamics on
the above decays, as interaction of the scalar mesons is intimately
connected with pseudoscalar loops.

The predictive possibilities of the model are illustrated in
Fig.~\ref{fig:1}. The observables shown in the figure are the invariant
mass distributions for the neutral pseudoscalar meson pairs in the $\phi$
radiative decays. We do not make a numerical fit to the data and just
illustrate flexibility of the model. The masses of the scalar mesons in
the analysis are taken $m_{f_0}=980.0$\,MeV and $m_{a_0}=984.7$\,MeV. The
parameters are chosen to reproduce the form of the distribution in the
vicinity of the peak (see the dotted line). Such a choice, $c_m =
-38.32$~MeV, $c_d = -9.47$~MeV and $\theta=-11.62^o$, also preserves the
reasonable agreement with the decay widths, see Table~\ref{tab:1}. The
table also shows our previous results~\cite{Ivashyn:2007yy}, which were
based on analysis of the average width values given by PDG~\cite{pdg}. A
comparison with predictions of other
models~\cite{Kalashnikova:2005zz,Black:2002ek,Nagahiro:2008mn} is
presented in Table~\ref{tab:1} as well. We also update the predictions
for the $S\to \gamma V$ because of the growing interest to these
decays~\cite{Ivashyn:2007yy,Kalashnikova:2005zz,Black:2002ek,Nagahiro:2008mn}.
One can see that the current approach describes the hadronic decays more
accurately than the two-photon decays.

\begin{table}
\begin{tabular}{crrrrrrr}
\hline
   \tablehead{1}{c}{b}{observable}
  & \tablehead{1}{r}{b}{estim.}
  & \tablehead{1}{r}{b}{\cite{Ivashyn:2007yy}\tablenote{an asterisk marks the input values}}
  & \tablehead{1}{r}{b}{~\cite{Kalashnikova:2005zz}\tablenote{the kaon loop model estimates
  are selected from Ref.~\cite{Kalashnikova:2005zz}}}
  & \tablehead{1}{r}{b}{model ``a'' \\Ref.~\cite{Black:2002ek}}
  & \tablehead{1}{r}{b}{model ``b'' \\Ref.~\cite{Black:2002ek}}
  & \tablehead{1}{r}{b}{\cite{Nagahiro:2008mn}\tablenote{the model of Ref.~\cite{Nagahiro:2008mn}
  accounts also for vector mesons in the loops}
  }
  & \tablehead{1}{r}{b}{PDG~\cite{pdg}}
\\
\hline
$\frac{\Gamma_{\phi \to \gamma a_0}}{\Gamma_{\phi}}\times 10^{-4} $
& $0.63$
& $1.7$
& $1.4 $
& ---
& ---
& ---
& $0.76 \pm 0.06$
\\
$\frac{\Gamma_{\phi \to \gamma f_0}}{\Gamma_{\phi}} \times 10^{-4}$
& $3.16$
& $4.4^*$
& $1.4$
& $4.92\pm 0.07$
& $4.92\pm 0.07$
& ---
& $4.40 \pm 0.21$
\\
$\frac{\Gamma_{\phi \to \gamma f_0}}{\Gamma_{\phi \to \gamma a_0}}$
& $4.99$
& $2.6$
& $1$
& $0.26 \pm 0.06$
& $0.46 \pm 0.09$
& ---
& $6.1 \pm 0.6$
\\
\noalign{\smallskip}\hline\noalign{\smallskip}
$ \Gamma_{a_0 \to \pi\eta}$,~MeV
& $21.11$
& $14.2$
& ---
& ---
& ---
& ---
& ---
\\
$\Gamma_{f_0 \to \pi\pi}$,~MeV
& $54.56$
& $41.8$
& ---
& ---
& ---
& ---
& $34.2 {}^{+ 22.7}_{- 14.3}$
\\
\noalign{\smallskip}\hline\noalign{\smallskip}
$\Gamma_{a_0 \to \gamma\gamma}$,~keV
& $0.16$
& $0.30^*$
& $0.24$
& $0.28\pm 0.09$$^*$ & $0.28\pm 0.09$$^*$
& ---
& $0.30 \pm {0.10}$
\\
$\Gamma_{f_0 \to \gamma\gamma}$,~keV
& $0.13$
& $0.31^*$
& $0.24$
& $0.39\pm 0.13$$^*$ & $0.39\pm 0.13$$^*$
& ---
& $0.31 {}^{+0.08}_{-0.11}$
\\
\noalign{\smallskip}\hline\noalign{\smallskip}
$\Gamma_{a_0 \to \gamma \rho}$,~keV
& $5$
& $9.1$
& $3.4 $
& $3.0 \pm 1.0$ & $3.0 \pm 1.0$
& $11\pm4$
& ---
\\
$\Gamma_{f_0 \to \gamma \rho}$,~keV & $4.1$ & $9.6$\tablenote{ both pion
and kaon loops are included in the current model}
& $3.4 $
& $4.2\pm1.1$
& $19 \pm 5$ & $3.3\pm 2.0$
& ---
\\
$\Gamma_{a_0 \to \gamma \omega}$,~keV
& $4.8$
& $8.7$
& $3.4$
& $641 \pm 87$ & $641 \pm 87$
& $31\pm13$
& ---
\\
$\Gamma_{f_0 \to \gamma \omega}$,~keV
& $8.4$
& $15.0$
& $3.4$
& $4.3\pm1.3$
& $126 \pm 20$ & $88\pm 17$
& ---
\\
\hline
\end{tabular}
\caption{The width estimates and comparison with other models}
\label{tab:1}
\end{table}

We should notice that the other resonance contributions, which interfere
with the scalar meson mechanism, are quite complicated, even in case of
the neutral particles in the final state (e.g., there are
$\phi\to\rho^0\pi^0\to\pi^0\pi^0\gamma$,
$\phi\to\omega\eta\to\eta\pi^0\gamma$, $\phi
\to\rho^0\pi^0\to\eta\pi^0\gamma$ and other contributions). In this
connection we note that a problem of the distribution data analysis is
discussed in Ref.~\cite{Bugg:2006sr}. It is a challenge to calculate
consistently the interfering contributions in framework of $R\chi T$ and
build a unified analysis of the available data.

The model used here for the mass distributions in the $\phi$ decays can
be applied to a detailed study of other radiative decays involving the
light scalar mesons. Among the most interesting ones we should mention
the $S\to \gamma V$ decays currently studied experimentally in
J\"ulich~\cite{Adam:2004ch,Marcus:this}. An option to apply the model for
the above decays and $\phi(1020)\to \pi^+\pi^-\gamma$ data is to include it in a Monte
Carlo generator for the Dalitz plot or mass distribution analysis.


\begin{theacknowledgments}
A.K. acknowledges support by the INTAS grant 05-1000008-8328
``Higher order effects in $e^+ e^-$ annihilation and muon
anomalous magnetic moment''. S.I. was partly supported by
National Academy of Science of Ukraine, grant
for young researchers~n.50-2007 and Joint NASU-RFFR scientific 
project N~01/50-2008.
\end{theacknowledgments}

\end{document}